\documentclass[]{article}

\usepackage{psfig}
\usepackage[sort&compress]{natbib}
\usepackage{amsmath}
\usepackage{bm}
\usepackage{epsfig}

\newcommand{\ket}[1]{\left. \vrule \, #1 \right\rangle}
\newcommand{\braket}[3]{\left\langle #1\, \vrule \, #2 \, \vrule \, #3 \right\rangle}
\newcommand{\scalprod}[2]{\left\langle #1\, \vrule \, #2 \right\rangle}

\newcommand{\vect}[1]{\mathbf{#1}}

\newcommand{\uvect}[1]{\hat{\mathbf{#1}}}
\newcommand{\uvecti}[1]{\hat{\boldsymbol{#1}}}
\newcommand{\muB}{\mu_{\textrm{B}}}
\newcommand{\degree}{^\circ}

\begin{document}

\bibliographystyle{unsrt}


\title{\bf Exact solution of the Zeeman effect \\ in single-electron systems}
\author{A. Blom\footnote{Email address: Anders.Blom@teorfys.lu.se}\\[+3mm]
Division of Solid State Theory, Department of Physics, \\ Lund University, S-223 62 Lund, Sweden}

\maketitle

\begin{abstract}
Contrary to popular belief, the Zeeman effect can be treated exactly in single-electron systems, for arbitrary magnetic
field strengths, as long as the term quadratic in the magnetic field can be ignored. These formulas were actually
derived already around 1927 by Darwin, using the classical picture of angular momentum, and presented in their proper
quantum-mechanical form in 1933 by Bethe, although without any proof. The expressions have since been more or less lost
from the literature; instead, the conventional treatment nowadays is to present only the approximations for weak and
strong fields, respectively. However, in fusion research and other plasma physics applications, the magnetic fields
applied to control the shape and position of the plasma span the entire region from weak to strong fields, and there is
a need for a unified treatment. In this paper we present the detailed quantum-mechanical derivation of the exact
eigenenergies and eigenstates of hydrogen-like atoms and ions in a static magnetic field. Notably, these formulas are
not much more complicated than the better-known approximations. Moreover, the derivation allows the value of the
electron spin gyromagnetic ratio $g_s$ to be different from 2. For completeness, we then review the details of dipole
transitions between two hydrogenic levels, and calculate the corresponding Zeeman spectrum. The various approximations
made in the derivation are also discussed in details.

\end{abstract}

\newpage


\section{Introduction}

At the end of the 19th century, when Maxwell's theory of electromagnetism had been established, it was known that
electromagnetic radiation is produced by oscillating electric charges. The charges that produce light were however
still a mystery. At the same time, it was generally believed that an electric current was made up of charged particles.
Lorentz suggested that atoms might also consist of charged particles, and that the oscillations of these particles
inside the atoms might be the source of light. If this were true, then a magnetic field ought to have an effect on the
oscillations and therefore on the wavelength of the light thus produced. In 1896 Zeeman, a pupil of Lorentz,
demonstrated this phenomenon~\cite{Zeeman}, known as the Zeeman effect, and in 1902 they were awarded the Nobel Prize.

Lorentz was thus one of the first to predict the existence of the electron, which, within a year later, was discovered
by Thomson, at least as a free particle. This, in combination with the results obtained by Planck and Einstein
regarding the nature of black-body radiation and the photo-electric effect, led to the subsequent emergence of Bohr's
atomic theory.

A full understanding of the Zeeman effect can only be obtained from quantum mechanics, but Lorentz was nevertheless
soon after the initial discovery by Zeeman able to produce a simple classical theory which predicted certain aspects of
the polarization of the different spectral components. These were later verified experimentally by Zeeman. Soon after
the discovery of the so-called anomalous Zeeman effect (the weak-field limit), around 1905, came the development of the
Land{\'e} vector model of the atom. Using this model, with its semi-empirical rules, basically all aspects of the
observed Zeeman spectra could be explained and predicted~\cite{X}. This was however most fortuitous, since the
classical Land{\'e} $g$-factor happens to be identical to its quantum-mechanical counterpart.

Based on the Land{\'e} vector model of angular momentum, in combination with the wave-mechanics of Schr{\"o}dinger,
Darwin~\cite{Darwin} was in 1927 able to solve "the problem of a spinning electrified body moving in a central orbit in
a magnetic field [\ldots] by the method of the wave mechanics in spherical harmonics". The formalism presented there
bears very little resemblance to how the problem is formulated nowadays, but the treatment is actually equivalent to an
exact solution of the Schr{\"o}dinger equation for the Zeeman effect.

It is certainly amusing to read Darwin's statement that "The proper attack on this problem would undoubtedly be by way
of the recent work of Heisenberg [\ldots] but this theory is still in the making, so it has not been practicable to
apply it here"; the article goes on to say that the work of Heisenberg "could give all the results of this paper; but
it would have been harder to follow because the matrix methods are not so easy for most readers as are spherical
harmonics."

Today the situation is rather the opposite, and we shall therefore in this paper present the exact eigenvalues and
eigenvectors for the Zeeman effect in single-electron ions (i.e.\ hydrogenic systems), using the matrix formulation of
quantum mechanics. These expressions are not new; they appeared as early as 1933, due to Bethe~\cite{bethe}, and they
have also been published in the well-known book by Bethe and Salpeter~\cite{BS}, where several aspects of the
fundamental properties of the Zeeman effect, as well as certain approximations made in the treatment, are discussed in
depth.

However, in neither of the above quoted works is given any indication -- or reference -- of how to derive these exact
formulas. Due to this, perhaps, the presentation of the Zeeman effect found in modern text-books on atomic physics (see
e.g.\ Ref.~\citealp{BJ}, Section 5.2), is limited to the discussion of the weak and strong field limits, known
respectively as the anomalous and normal Zeeman effect, and perhaps some particular intermediate cases, where the
Zeeman effect is also known as the Paschen--Back effect. (By strong and weak fields, we mean that the energy
contribution from the magnetic field is large or small compared to the spin--orbit interaction.)

For many-electron atoms, this is naturally the proper attack on the problem, since an exact solution is unattainable.
However, for the particular case of hydrogenic atoms and ions, the exact treatment is not much more complicated; in
fact, the resulting expressions for the eigenvalues and eigenvectors are quite simple. The spectral properties of
single-electron ions have recently received rekindled interest due to a variety of applications within plasma
spectroscopy (for a review, see Ref.~\citealp{Islerreview} and references therein). In e.g.\ fusion research, strong
magnetic fields are used to control the shape and position of the plasma. A detailed understanding of the influence of
these fields on the spectral lines is crucial for a correct interpretation of the observed line-widths, which are
employed for temperature measurements~\cite{hey,ppcf}, among other things. Moreover, the magnetic field strengths
employed often span the entire scale from weak to strong fields, and there is therefore a clear need for a unified
treatment of the Zeeman effect, valid for all fields.

The structure of this paper is as follows: in Section~\ref{sec:exact} we will show how to solve the Schr{\"o}dinger
equation exactly for a hydrogenic system in a magnetic field, under the assumption that the quadratic term in the
magnetic field strength can be ignored. We then show in Section~\ref{sec:strongweak} how expansions of the exact
expressions reproduce the traditional results of weak and strong fields. As a bonus, a quantitative expression is found
for determining the validity of these approximations. States with zero angular momentum must be considered separately,
and actually correspond to an extreme case of the strong-field approximation, as discussed in
Section~\ref{sec:s-states}. Section~\ref{sec:HTrans} contains a review of the dipole transition matrix elements and
selection rules, and finally we discuss in Section~\ref{sec:Approxs} the approximations made during the derivations.


\section{Exact solution of the Schr{\"o}dinger equation} \label{sec:exact}

As pointed out above, the traditional treatment of the Zeeman effect in most text-books is limited to the cases where
the influence of the magnetic field can be separated from the effects of the spin--orbit interaction. In the weak-field
limit, the magnetic interaction is introduced as a perturbation to the spin--orbit eigenstates, and vice-versa in the
strong-field case. In this picture, all perturbations are already diagonal and the formalism is trivial.

In the general case, to be presented here, these two effects must however be treated simultaneously. Our method will be
to diagonalize the combined perturbation in a basis set consisting of the solutions of the field-free, non-relativistic
Schr{\"o}dinger equation for a single-electron ion. Depending on the orbital quantum number $\ell$ of the states
considered, the matrices will be of various dimensions. A significant simplification will however emerge from the
algebra, and the final energy matrix will separate into a number of $2 \times 2$ sub-matrices which are easily
diagonalized. The case $\ell=1$ is treated in some detail in Ref.~\citealp{Fano}, from where the idea for the method
has been adopted and further developed.


\subsection{The unperturbed system} \label{sec:FieldFreeSE}

The non-relativistic Schr{\"o}dinger equation for a hydrogenic system in vacuum, ignoring for now the spin--orbit
interaction, is
\begin{equation}\label{eq:FieldFreeSE}
\left[ -\frac{\hbar^2}{2\mu} \nabla^2 - \frac{Ze^2}{4 \pi \epsilon_0 r} \right] \psi(r) = H_0\psi(r) = E_0\psi(r),
\end{equation}
where $\mu = m_{\textrm{e}} M/(m_{\textrm{e}}+M)$ is the reduced electron mass ($m_{\textrm{e}}$ and $M$ are the
electron and nuclear masses, respectively) and $Z$ the nuclear charge. Because of the spherical symmetry, the
wavefunction $\psi$ can be written as a product
\begin{equation}
\ket{n \ell m_\ell} = \psi_{n \ell m_\ell} (r,\theta,\varphi) = R_{n\ell}(r) Y_{\ell m}(\theta,\varphi)
\end{equation}
of an angular part, given by the spherical harmonics $Y_{\ell m}(\theta,\varphi)$, and a radial part $R_{n\ell}(r)$.

The energy of each eigenstate $\ket{n\ell m_\ell}$ will depend only on the principal quantum number $n$ as
\begin{equation}\label{eq:E0}
E_0 = -\frac{\mu c^2}{2} \frac{(Z\alpha)^2}{n^2},
\end{equation}
where $\alpha = e^2/4\pi\epsilon_0\hbar c \approx 1/137.036$ is the fine-structure constant. Introducing relativistic
corrections for the momentum operator adds another term
\begin{equation}\label{eq:RelCorr}
\Delta E_{\textrm{rel}} = \frac{1}{2} m_{\textrm{e}} c^2 \frac{(Z\alpha)^2}{n^4} \left[ \frac{3}{4}- \frac{n}{\ell+1/2}
\right]
\end{equation}
to the energy, breaking the degeneracy in the $\ell$ quantum number. The degeneracy in the $m$ quantum number is only
broken through an external field such as the magnetic field about to be studied.

It is not entirely obvious which value of the electron mass to use in Eq.~\eqref{eq:RelCorr}, but since the
perturbation in itself is a small correction, it will make an insignificant difference whether we use the electron mass
$m_{\textrm{e}}$ or the reduced mass $\mu$~\cite[footnote on p.~196]{BJ}. Therefore, we shall in what follows always
use the standard electron mass in perturbation theory.


\subsection{Interaction with an external field} \label{sec:ExternalField}

The interaction between an atom and an external electromagnetic field is introduced through the vector potential
$\vect{A}$. From Maxwell's equation $\nabla\cdot\vect{B}=0$ (there are, probably, no magnetic monopoles), it is clear
that one can choose to represent any magnetic field $\vect{B}$ as $\vect{B}=\nabla \times \vect{A}$. The choice of
$\vect{A}$ is not unique, but one can generally write $\vect{A} = ( \vect{B} \times \vect{r} )/2 + \nabla\Psi$, where
$\Psi$ is an arbitrary scalar potential. Note that over atomic distances, the macroscopic magnetic field can be assumed
to be constant in magnitude and direction.

The non-relativistic Schr{\"o}dinger equation (relativistic corrections and other approximations are discussed in
Section~\ref{sec:Approxs}) for a spin-less electron in an external magnetic field can be derived from the Klein--Gordon
equation~\cite{BS}, and is
\begin{equation}
\left[ -\frac{\hbar^2}{2\mu} \nabla^2 - \frac{Ze^2}{4\pi\epsilon_0 r} - \frac{i\hbar e}{m_{\textrm{e}}} \vect{A} \cdot
\nabla + \frac{e^2}{2m_{\textrm{e}}} \vect{A}^2 \right] \psi (\vect{r}) = E\psi(\vect{r}) ,
\end{equation}
where we also have used the freedom of gauge invariance to set $\nabla \cdot \vect{A} = 0$ (the Coulomb gauge), which
means that the scalar field $\Psi$ must be a constant, which we arbitrarily may choose as zero.

The term linear in $\vect{A}$ can be written
\begin{equation*}
-\frac{i\hbar e}{m_{\textrm{e}}} \vect{A} \cdot \nabla = \frac{e}{2m_{\textrm{e}}} \vect{B} \cdot
\vect{L},
\end{equation*}
where $\vect{L} = -i \hbar (\vect{r} \times \nabla )$ is the angular momentum operator.
Similarly, the quadratic term becomes
\begin{equation*}
\frac{e^2}{2m_{\textrm{e}}} \vect{A}^2 = \frac{e^2}{2m_{\textrm{e}}}\left( \vect{B} \times \vect{r} \right)^2.
\end{equation*}
Comparing the magnitudes of the linear and quadratic terms, one finds~\cite{BJ} that their ratio is of the order $B
\cdot 10^{-6}$/Tesla for atomic systems. Except for extreme cases, such as neutron stars, the quadratic term is hence
completely negligible, which simplifies the mathematics a great deal. (When included, the quadratic term will lead to
diamagnetism~\cite{BS}.) For very strong magnetic fields, when the spin--orbit interaction can be neglected altogether
but the quadratic term not, the energy levels can be found by different methods~\cite{simola}.

So far we have used a semi-classical picture of the electron by ignoring the spin, which however contributes an
additional, intrinsic, angular momentum. To the magnetic moment of the electron from the classical angular momentum
\begin{equation}
\vect{M}_{\textrm{L}} = -\frac{e}{2m_{\textrm{e}}}\vect{L} = -\muB \vect{L} / \hbar ,
\end{equation}
where $\muB$ is the Bohr magneton, we should therefore add a contribution from the spin
\begin{equation}
\vect{M}_{\textrm{S}} = -g_s\frac{e}{2m_{\textrm{e}}}\vect{S} = -g_s\muB \vect{S} / \hbar.
\end{equation}
According to the original Dirac theory, the magnetic moment of the electron due to its intrinsic spin is determined by
$g_s\equiv 2$. However, using quantum electrodynamics (QED), one finds that the value of the spin gyromagnetic ratio
$g_s$ is not exactly integer, but
\begin{equation}
\frac{g_s}{2} = 1 + \frac{\alpha}{2\pi} + \frac{0.328\alpha^2}{\pi^2}+\hdots \approx 1.0011595 .
\end{equation}
This correction is called the anomalous magnetic moment of the electron. Since it poses no problems to retain a general
value of $g_s$ in our calculations, we will do so. Moreover, this allows the formalism to be extended (with proper
modifications, accounting for the relative dielectric constant and the electron effective mass) to e.g.\ solids, where
the values of the gyromagnetic ratio may differ significantly from 2, and even be negative~\cite{callaway}.

The energy of a magnetic dipole -- which the electron effectively becomes due to its magnetic moment -- in a magnetic
field is
\begin{equation}\label{eq:Hmagn}
H'_{\textrm{magn}} = -(\vect{M}_{\textrm{L}}+\vect{M}_{\textrm{S}}) \cdot \vect{B} = \frac{\muB}{\hbar} \left(
\vect{L}+g_s\vect{S} \right)\cdot \vect{B} ,
\end{equation}
and we now also introduce the energy from the spin--orbit interaction
\begin{equation}
H'_{\textrm{so}} = \xi(r) \left( \vect{L} \cdot \vect{S} \right) ,\end{equation} where for the one-electron
central-field case
\begin{equation}
\xi(r) = \frac{1}{2m_{\textrm{e}}^2 c^2} \frac{Ze^2}{4\pi\epsilon_0} \frac{1}{r^3} .\end{equation}

In the field-free Schr{\"o}dinger equation, Eq.~\eqref{eq:FieldFreeSE}, the geometry is spherically symmetric, whereas
the magnetic field defines a particular direction of symmetry. We make this direction the $z$-axis of our system, which
reduces the scalar product in Eq.~\eqref{eq:Hmagn} to $\vect{B} \cdot \left( \vect{L} + g_s\vect{S} \right) = B \left(
L_z + g_sS_z \right)$, where $B = |\vect{B}|$.

The total non-relativistic Hamiltonian for the electron in a hydrogen-like atom in an external magnetic field thus
becomes
\begin{equation} \label{eq:TotalHamiltonian}
\begin{split}
H &= H_0 + H'_{\textrm{so}} + H'_{\textrm{magn}} \\
&=  -\frac{\hbar^2}{2\mu} \nabla^2 - \frac{Ze^2}{4 \pi \epsilon_0 r} + \frac{1}{2m_{\textrm{e}}^2 c^2}
\frac{Ze^2}{4\pi\epsilon_0} \frac{1}{r^3} \left( \vect{L} \cdot \vect{S} \right) + \frac{\muB B}{\hbar} \left(
L_z+g_sS_z\right) .
\end{split}
\end{equation}
We assume that the relativistic correction $\Delta E_{\textrm{rel}}$ is still given by Eq.~\eqref{eq:RelCorr} in the
presence of the external field.


\subsection{Perturbed energies and eigenstates}

To diagonalize the perturbation
\begin{equation}
H' = H'_{\textrm{so}} + H'_{\textrm{magn}}
\end{equation}
we first need to introduce a proper set of basis states. This is actually not trivial~\cite[Section 45$\beta$]{BS}, but
to simplify matters we shall use the Pauli approximation, in which each state is a two-component spinor and the spin
operators are described by the Pauli spin matrices.

We can therefore use as basis the eigenstates of the unperturbed Hamiltonian $H_0$, arranged in a proper spinor
notation. In the unperturbed system, the good quantum numbers are, in addition to $n$ and $\ell$, also $m_\ell$ and
$m_s$, although the eigenstates are degenerate in the $m$:s. In the presence of the external field, the individual
$m$:s are no longer good quantum numbers, but only their sum $m=m_\ell+m_s$.

We can therefore anticipate that each new eigenstates will be a linear combination of exactly two unperturbed
eigenstates, corresponding to $m_s=\pm \frac{1}{2}$ and $m_\ell$ chosen accordingly. But since the spin--orbit
interaction is not diagonal in the unperturbed eigenstates, we need to carry out the derivation using the full
expansion in the unperturbed eigenstates
\begin{equation}
\ket{n \ell m} = \sum_{m_\ell} \sum_{m_s} \ket{n\ell m_\ell m_s} \scalprod{n\ell m_\ell m_s}{n \ell m} ,\end{equation}
where $m_s$ and $m_\ell$ take on all allowed values. To avoid unnecessary notation, the indices $n$ and $\ell$ will be
suppressed most of the time, as all calculations are performed for fixed values of these quantum numbers. In all
matrices below, the basis states are assumed to be arranged in the following order:
\begin{equation*}
(m_\ell,m_s) = (\ell,\uparrow), \; (\ell,\downarrow), \; (\ell-1,\uparrow), \; (\ell-1,\downarrow), \; \ldots , \;
(-\ell,\uparrow), \; (-\ell,\downarrow)
\end{equation*}
where up and down arrows mean $m_s=\pm \frac{1}{2}$ respectively.

The contribution to the energy due to $H'$ is
\begin{equation}\label{eq:Delta_E_Perturb}
\begin{split}
\Delta E_{n\ell m} (B) &= \braket{n\ell m}{H'}{n\ell m} \\[+3mm]
&= W \sum_{\substack{m_\ell,m_s \\ m'_\ell,m'_s}} \scalprod{n\ell m}{m_\ell m_s}  \braket{m_l
m_s}{\frac{2\vect{L}\cdot\vect{S}}{\hbar^2} + \frac{L_z+g_sS_z}{\hbar}\beta} {m'_\ell m'_s} \scalprod{m'_\ell m'_s}{n
\ell m} ,\end{split}
\end{equation}
where we have defined
\begin{equation}
\beta = B/B_0,
\end{equation}
with $B_0 = W/\muB$ where
\begin{equation} \label{eq:WB0}
W = \frac{\hbar^2}{2} \big\langle \xi(r) \big\rangle = \muB^2 \frac{Z}{4\pi\epsilon_0 c^2} \left\langle \frac{1}{r^3}
\right\rangle_{n\ell}.
\end{equation}
The expectation value $W$ is not defined for $\ell=0$, since then there is no spin--orbit interaction, but instead an
energy contribution named the Darwin term. We will consider that case separately in Section~\ref{sec:s-states}, and may
assume $\ell \neq 0$ in what follows. One can then show~\cite{BJ} that for the hydrogenic eigenfunctions,
\begin{equation}
\left\langle \frac{1}{r^3} \right\rangle_{n\ell} = \frac{Z^3}{a_0^3 n^3 \ell(\ell+1/2)(\ell+1)},
\end{equation}
and thus
\begin{equation} \label{eq:Wnl}
W = \frac{m_{\textrm{e}} c^2}{4} \frac{(Z\alpha)^4}{n^3 \ell (\ell+1/2)(\ell+1)}.
\end{equation}

The magnetic field term,
\begin{equation} \label{eq:magnmatrix}
\braket{m_\ell m_s}{\frac{L_z+g_sS_z}{\hbar}}{m'_\ell m'_s} = \left( m_\ell+g_s m_s \right) \delta_{m_\ell
m'_\ell}\delta_{m_s m'_s},
\end{equation}
is already diagonal in the present basis, but the spin--orbit interaction $\vect{L} \cdot \vect{S}= L_x S_x+L_y S_y+L_z
S_z$ is not. Since the orbital angular momentum operator $\vect{L}$ and the spin angular moment operator $\vect{S}$ act
on separate vector spaces, spin and position space respectively, they trivially commute. Therefore, we can separate
space and spin, so that
\begin{equation}\label{eq:SepLS}
\braket{m_\ell m_s}{L_i S_i}{m'_\ell m'_s} = \braket{m_\ell }{L_i}{m'_\ell } \braket{m_s}{S_i}{m'_s} \qquad (i=x,y,z).
\end{equation}
This expression corresponds to an outer or Kronecker product between the space and spin matrices. This is defined as
\begin{equation}\label{eq:MatrixProduct}
\begin{split}
\vect{A} \otimes \vect{B} &\equiv
\begin{pmatrix}
a_{11} & a_{12} & a_{13} \\
a_{21} & a_{22} & a_{23} \\
a_{31} & a_{32} & a_{33}
\end{pmatrix}
\otimes
\begin{pmatrix}
b_{11} & b_{12} \\
b_{21} & b_{22} \\
\end{pmatrix}
=
\begin{pmatrix}
a_{11}b_{11} & a_{11}b_{12} &\vdots& a_{12}b_{11} & a_{12}b_{12} &\vdots& a_{13}b_{11} & a_{13}b_{12}
\\[-1.5mm]
a_{11}b_{21} & a_{11}b_{22} &\vdots& a_{12}b_{21} & a_{12}b_{22} &\vdots& a_{13}b_{21} & a_{13}b_{22}
\\[-1.5mm]
\hdotsfor{8} \\[-2mm]
a_{21}b_{11} & a_{21}b_{12} &\vdots& a_{22}b_{11} & a_{22}b_{12} &\vdots& a_{23}b_{11} & a_{23}b_{12}
\\[-1.5mm]
a_{21}b_{21} & a_{21}b_{22} &\vdots& a_{22}b_{21} & a_{22}b_{22} &\vdots& a_{23}b_{21} & a_{23}b_{22}
\\[-1.5mm]
\hdotsfor{8} \\[-2mm]
a_{31}b_{11} & a_{31}b_{12} &\vdots& a_{32}b_{11} & a_{32}b_{12} &\vdots& a_{33}b_{11} & a_{33}b_{12}
\\[-1.5mm]
a_{31}b_{21} & a_{31}b_{22} &\vdots& a_{32}b_{21} & a_{32}b_{22} &\vdots& a_{33}b_{21} & a_{33}b_{22}
\end{pmatrix}
,\end{split}
\end{equation}
from which generalizations to larger matrices should be obvious. Each element in the product matrix is a product of one
element in the first and one in the second matrix, spanning all possible combinations. Within each marked sub-matrix,
one finds the matrix $\vect{B}$ multiplied with the subsequent elements in $\vect{A}$.

The required matrix elements for the spin part are immediately given by the Pauli spin matrices
\begin{equation}\label{eq:PauliMatrices}
\begin{split}
\sigma_x &= \braket{m_s}{S_x}{m'_s} = \frac{\hbar}{2}
\begin{pmatrix} 0&1 \\ 1&0 \end{pmatrix} ,\\
\sigma_y &= \braket{m_s}{S_y}{m'_s} = \frac{\hbar}{2}
\begin{pmatrix} 0&-i \\ i&0 \end{pmatrix} ,\\
\sigma_z &= \braket{m_s}{S_z}{m'_s} = \frac{\hbar}{2} \begin{pmatrix} 1&0 \\ 0&-1
\end{pmatrix} ,
\end{split}
\end{equation}
whereas for the space part we make use of the ladder operators
\begin{equation}\label{eq:LadderOps}
L_\pm = L_x \pm i L_y \quad \Leftrightarrow \quad \left\{ \begin{array}{l}
L_x = ( L_+ + L_- )/2, \\[+1mm]
L_y = ( L_+ - L_- )/2i.
\end{array} \right.
\end{equation}
whose action on the unperturbed eigenstates is defined through the relations (with conventional phase choices)
\begin{equation}\label{eq:Ladder_lm}
L_\pm \ket{\ell,m_\ell} = \hbar \sqrt{\ell(\ell+1)-m_\ell(m_\ell\pm 1)} \ket{\ell,m_\ell\pm 1} .
\end{equation}

To write down the matrices corresponding to the operators $L_x$ and $L_y$ is somewhat tricky, since the sizes of the
matrices depend on the value of $\ell$; the dimensions will be $2\ell+1$, i.e.\ the number of possible $m$-values. The
$z$-component is however diagonal, since the operators $L_z$ and $S_z$ correspond to the good quantum numbers $m_\ell$
and $m_s$ of the basis eigenstates, so we can write it generally as
\begin{equation}\label{eq:LzSz}
\braket{m_\ell }{L_z}{m'_\ell } \braket{m_s}{S_z}{m'_s} = m_\ell m_s \hbar^2 \delta_{m_\ell m'_\ell} \delta_{m_s m'_s}.
\end{equation}
This will be the only contribution to the diagonal of the spin--orbit term, since the ladder operators $L_\pm$ (and
hence $L_x$ and $L_y$) do not couple states with the same $m_\ell$ or $m_s$.

Moreover, although each of the matrices $\braket{m_\ell}{L_x} {m_\ell'}$ and $\braket{m_\ell}{L_y} {m_\ell'}$ contain
several off-diagonal elements, an important simplification takes place when they are added together to form
$\braket{m_\ell m_s}{\vect{L}\cdot\vect{S}} {m_\ell' m_s'}$, and the matrix separates into a sequence of $2\times 2$
sub-matrices, plus the corner elements which are uncoupled. Clearly, this represents what was predicted earlier: only
states of equal $m=m_\ell+m_s$ are coupled.

Using these two observations, it becomes possible to present the structure of the spin--orbit matrix as
\addtocounter{equation}{1}
\begin{equation} \tag{\theequation} \label{eq:GeneralLS}
\braket{m_\ell m_s}{\frac{2\vect{L}\cdot\vect{S}}{\hbar^2}}{m'_\ell m'_s} = \begin{pmatrix}
\ell & \vdots & 0     & 0       & \vdots & 0        & 0 & \vdots &     \\[-2.5mm]
\hdotsfor{9}\\[-2.5mm]
0    & \vdots & -\ell & ?       & \vdots & 0        & 0 & \vdots &   \\[-1.5mm]
0    & \vdots & ?     & \ell-1 & \vdots & 0         & 0 & \vdots &    \\[-2.5mm]
\hdotsfor{9}\\[-2.5mm]
0    & \vdots & 0     & 0       & \vdots& -(\ell-1) & ? & \vdots &    \\[-1.5mm]
0    & \vdots & 0     & 0       & \vdots& ?         & \ell-2 & \vdots & \\[-2.5mm]
\hdotsfor{9}\\[-2.5mm]
     & \vdots &       & & \vdots&           &        & \vdots & \ddots
\end{pmatrix},
\end{equation}
where the so-far undetermined elements are denoted "?". The structure is emphasized by the dotted lines, which enclose
$2\times2$ areas in which all matrix elements correspond to given quantum numbers $m$ and $m'$. It should be apparent
that the non-zero matrix elements will be confined to the block-diagonal, and that in each such block the relation
$m=m'$ will hold.

The corner elements of the matrix are trivial, so we focus on the interior of the matrix. To find the unknown matrix
elements, we may use the same matrix but with a different division into sub-blocks:
\begin{equation} \label{eq:GeneralLS2} \tag{\theequation$'$}
\braket{m_\ell m_s}{\frac{2\vect{L}\cdot\vect{S}}{\hbar^2}}{m'_\ell m'_s} = \begin{pmatrix}
\ell & 0     & \vdots & 0       & 0         & \vdots & 0      \\[-1.5mm]
0    & -\ell & \vdots & \boxed{?}       & 0         & \vdots & 0      \\[-2.5mm]
\hdotsfor{8}\\[-2.5mm]
0    & ?     & \vdots & \ell-1 & 0          & \vdots & 0      \\[-1.5mm]
0    & 0     & \vdots & 0       & -(\ell-1) & \vdots & ?      \\[-2.5mm]
\hdotsfor{8}\\[-2.5mm]
0    & 0     & \vdots & 0       & ?         & \vdots & \ell-2 \\[-1.5mm]
     &        & \vdots &        &           & \vdots &        & \ddots
\end{pmatrix}
\end{equation}
This division corresponds exactly to the direct matrix product~\eqref{eq:MatrixProduct}, and we can therefore easily
trace the contributions to the unknown elements.

First consider the top right question mark (boxed) in Eq.~\eqref{eq:GeneralLS2}. In this area of the matrix,
$m'_\ell=\ell-1$ and $m_\ell=\ell$. These states are coupled by $L_+$ and we get for the sub-matrix
\begin{equation*}
\begin{split}
\frac{\hbar^2}{2} \begin{pmatrix} 0 & 0 \\ ? & 0 \end{pmatrix}
&= L_x \sigma_x + L_y \sigma_y \\
&= \braket{m_\ell=\ell}{L_+}{m'_\ell=\ell-1} \left( \frac{1}{2} \sigma_x + \frac{1}{2i} \sigma_y \right)  \\
&= \frac{\hbar^2}{2} \sqrt{\ell(\ell+1)-m'_\ell(m'_\ell+1)}
\begin{pmatrix} 0 & 0 \\ 1 & 0 \end{pmatrix}.
\end{split}
\end{equation*}
For the unknown element in the third row, second column, the only difference is that the involved states are coupled by
$L_-$ instead. The prime is only a book-keeping label; $m_\ell$ and $m'_\ell$ are interchangeable as long as $L_+$ is
replaced by $L_-$ and vice versa, and hence this second unknown element is exactly the same as the first one, with the
only change of removing the primes.

In the positions of the matrix where the question marks appear, we always have $m_s=1/2$ or $m_s'=1/2$. Therefore, on
these positions, $m_\ell=m-1/2$ (with or without primes). These observations are however not particular for this
sub-block of the matrix, but apply generally, and thus all non-zero off-diagonal elements can be written on the same
identical form $\sqrt{\ell(\ell+1)-(m-1/2)(m+1/2)}$, since furthermore $m=m'$ within each sub-block on the diagonal, as
noted above.

Finally we rewrite the diagonal elements, which have contributions from both the $L_zS_z$ matrix elements,
Eq.~\eqref{eq:LzSz}, and the magnetic interaction, Eq.~\eqref{eq:magnmatrix}:
\begin{equation*}
\begin{split}
\mathcal{D}(m_\ell,m_s) &= \braket{m_\ell
m_s}{\frac{2\vect{L}\cdot\vect{S}}{\hbar^2}+\beta\frac{L_z+g_sS_z}{\hbar}}{m_\ell m_s}
\\[+2mm]%
&= 2m_\ell m_s +\beta\left( m_\ell+g_s m_s \right)
\end{split}
\end{equation*}
When $m_s=\pm1/2$ we must have $m_\ell=m\mp1/2$ and thus
\begin{equation}
\mathcal{D}(m_\ell=m\mp1/2,m_s=\pm1/2) = \pm\left(m \mp \frac{1}{2}\right) +\beta\left[ m\pm \frac{g_s-1}{2} \right].
\end{equation}
Again, these expressions will be valid anywhere on the diagonal, and we have thus proven that all the $2\times2$ blocks
along the block diagonal can be written on a common form,
\begin{equation}\label{eq:FinalMatrix}
\begin{pmatrix}
-(m+1/2)+\beta[m-(g_s-1)/2] & \sqrt{\ell(\ell+1)-(m-1/2)(m+1/2)} \\[+3mm]
\sqrt{\ell(\ell+1)-(m-1/2)(m+1/2)} & m-1/2+\beta[m+(g_s-1)/2]
\end{pmatrix},
\end{equation}
valid for all allowed values of $m$, except of course the corner elements $|m|=\ell+1/2$. These are however trivial,
and give the eigenvalues (cf.\ Eq.~\eqref{eq:Delta_E_Perturb})
\begin{equation} \label{eq:SolutionMaxM} \Delta E_{\pm(\ell+1/2)} = W\Big[
\ell \pm \beta(\ell+g_s/2) \Big].
\end{equation}

To find the other solutions, we need only diagonalize the matrix~\eqref{eq:FinalMatrix}, which has eigenvalues
\begin{equation} \label{eq:FinalEnergy}
\Delta E_m^\pm = \frac{W}{2} \left[ 2m\beta-1 \pm \sqrt{\beta^2(g_s-1)^2+4m\beta(g_s-1)+(2\ell+1)^2} \right],
\end{equation}
which thus is the total energy shift due to the spin--orbit interaction and the magnetic field. The corresponding
eigenstates can be expressed in terms of the original $\ket{m_\ell,m_s}$ states as \addtocounter{equation}{1}
\begin{align} \label{eq:ZeemanEigenstates} \tag{\theequation a}
\ket{m}_+ &= \sin\varphi \ket{m+1/2,\downarrow} + \cos\varphi \ket{m-1/2,\uparrow},
\\[+2mm] \tag{\theequation b}
\ket{m}_- &= \cos\varphi \ket{m+1/2,\downarrow} - \sin\varphi \ket{m-1/2,\uparrow},
\end{align}
where the angle $\varphi$ is given by
\begin{equation}\label{eq:angle}
\varphi = \left| \frac{1}{2} \arctan \left( \frac{\sqrt{4\ell(\ell+1)-4m^2+1}}{2m+\beta(g_s-1)} \right) \right|.
\end{equation}
The absolute signs appear here since an explicit sign convention has been introduced in
Eq.~(\ref{eq:ZeemanEigenstates}b).

\begin{figure}[hbt]
\centering \epsfig{file=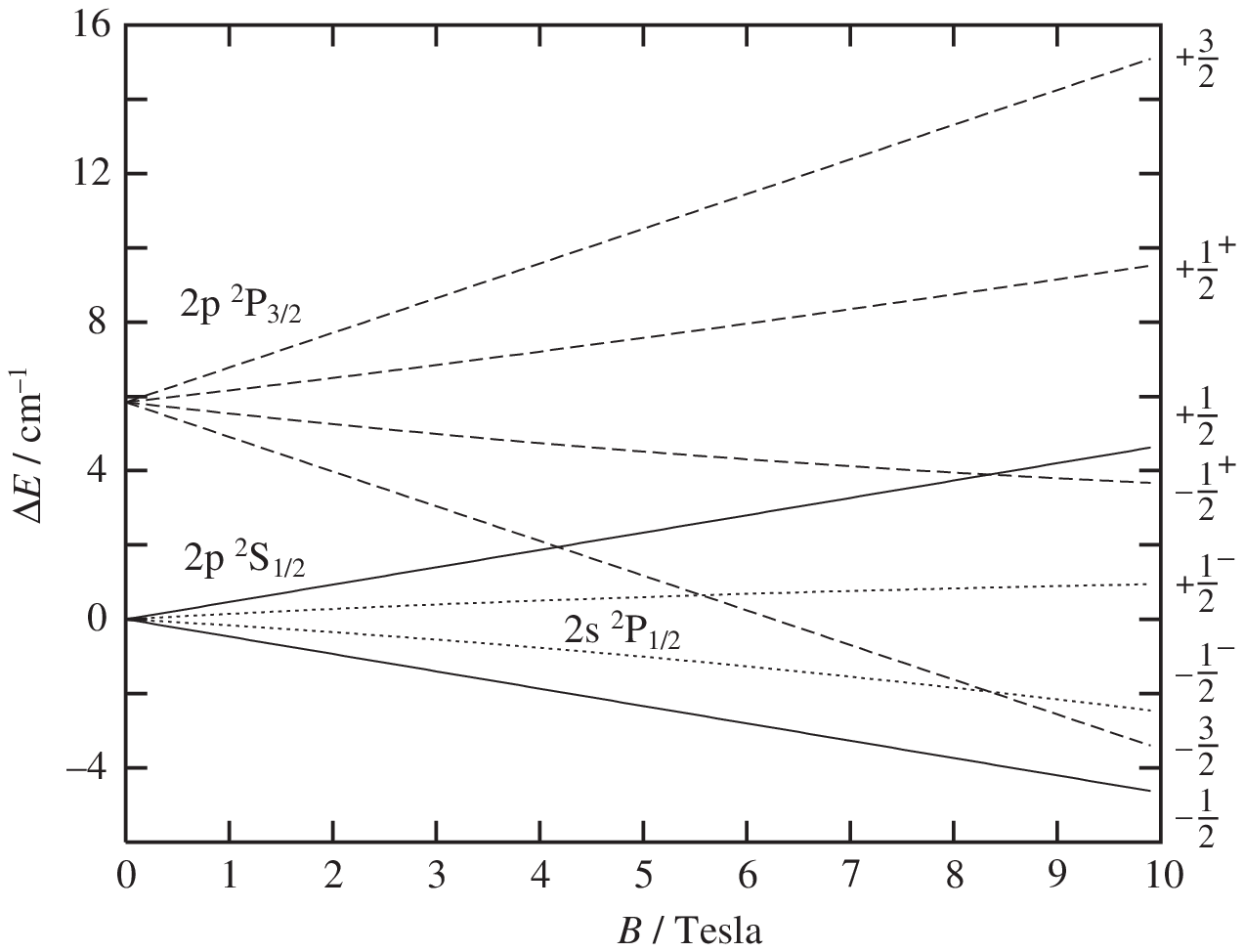, width=\linewidth} \caption{The total energy shift due to both spin--orbit
interaction and an applied magnetic field for the $n=2$ shell in HeII. The energy zero-point is chosen at
2s~$^2$S$_{1/2}$ at $B=0$. The states are marked by their quantum number $m$ with a following $\pm$ corresponding to
the notation in Eqs.~\eqref{eq:FinalEnergy} and (\ref{eq:ZeemanEigenstates}b), except for the states with
$|m|=\ell+1/2$ for which there only are two unique solutions, Eq.~\eqref{eq:SolutionMaxM}. For small magnetic fields,
the states are split linearly according to Eq.~\eqref{eq:WeakFieldExpansion}. Note that without any magnetic field, the
two levels 2s~$^2$S$_{1/2}$ and 2p~$^2$P$_{1/2}$ are degenerate (the small difference due to the Lamb shift is
ignored). Also see Fig.~\ref{fig:zeeB20}, where the field scale is extended.} \label{fig:zeeB2}
\end{figure}

\begin{figure}[hbt]
\centering \epsfig{file=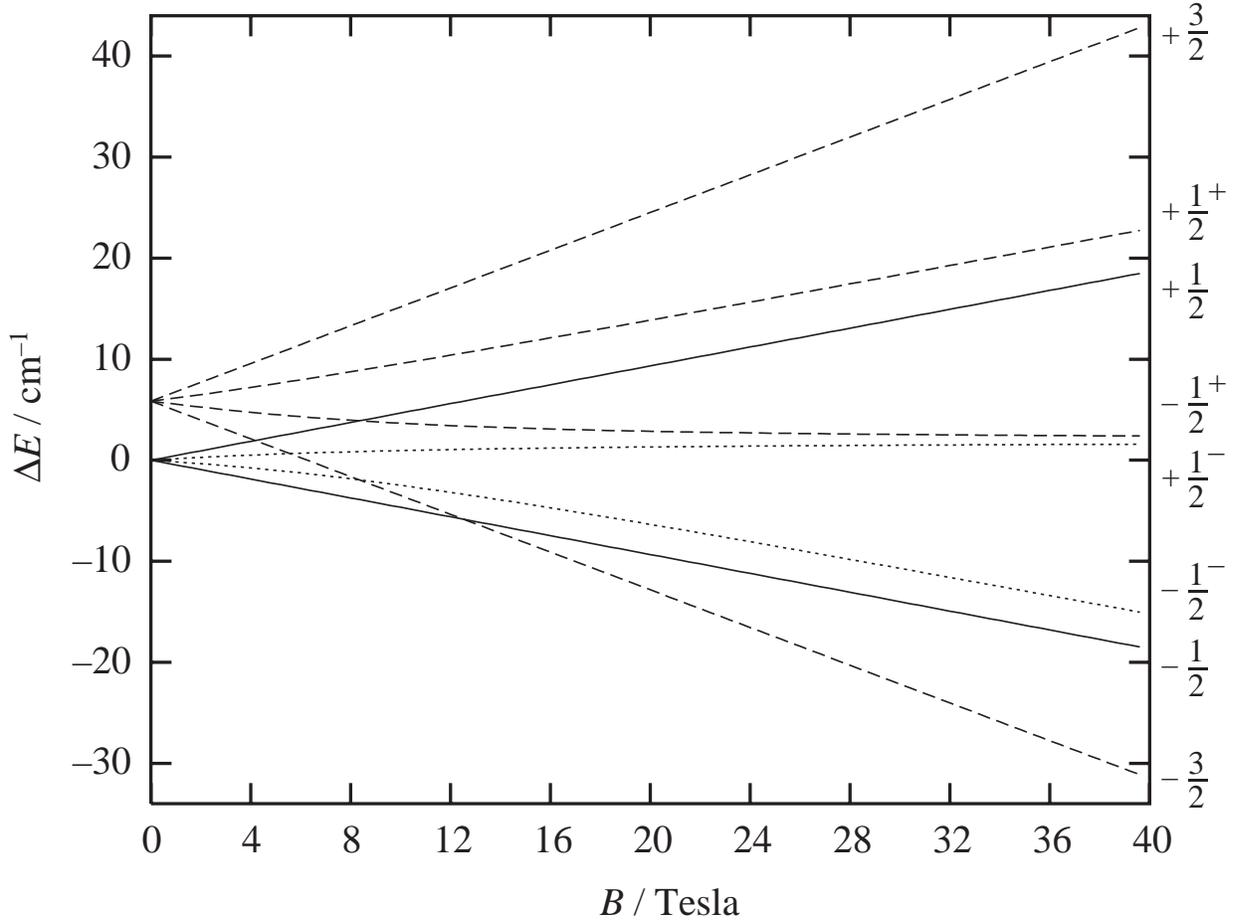, width=\linewidth} \caption{In this figure, the same system as in
Fig.~\ref{fig:zeeB2} is shown, but for stronger fields. As the magnetic field approaches high values, the pattern
converges into five states, according to Eq.~\eqref{eq:StrongFieldExpansion}. Since $m$ takes on the values $\pm 1/2,
\pm 3/2$, there are only five possible values of $m\pm 1/2$, namely $0,\pm 1,\pm 2$, and the degeneracy will obviously
be one for the $\pm 2$ cases (only reached from $m=\pm 3/2$), and two for the other, as is also seen in the figure.
Noteworthy is also the anti-crossing rule: even if some lines cross as $B$ increases, no lines with the same $m$-values
ever cross. This is an effect of level repulsion between states of equal quantum mechanical symmetry (cf.\
Ref.~\citealp{Cowan}, pp.~292--295 and Fig.~10-2 therein.)} \label{fig:zeeB20}
\end{figure}

Examples of the calculated Zeeman levels based on these results are shown in Figs.~\ref{fig:zeeB2} and
\ref{fig:zeeB20}.


\section{Strong and weak field expansions}\label{sec:strongweak}

The formalism as presented above makes no assumptions of the relative strength between the spin--orbit interaction and
the magnetic field. In the traditional derivation, however, the one of the two which is stronger is first treated as a
perturbation to the central-field Hamiltonian, which gives a set of new eigenfunctions. Subsequently, the other effect
is treated as a perturbation to this new system, using its eigenfunctions as a basis set.

In many practical cases the magnetic fields in question are either "weak" -- laboratory field strengths are typically
of the order 10--100~mT -- or "strong" (superconducting electromagnets or astrophysical objects), making the
traditional approach seemingly more appealing, since the algebra and the resulting expressions are somewhat simpler
(but only very little!).

It is therefore desirable to attempt to derive these limits from the general case, and also to obtain a quantitative
criterion for when these approximations are valid, which is generally brushed over in the traditional texts, where
"strong" or "weak" fields are merely assumed. For the weak-field case, it is necessary to assume $g_s=2$, and therefore
we will also assume this for strong fields, in order to facilitate the comparison with expressions in the literature.

As our starting point we will use the expression for $\Delta E_m^\pm$, Eq.~\eqref{eq:FinalEnergy}. The parameter
determining the relative strength of the magnetic interaction to the spin--orbit interaction is $\beta$. Assuming that
this quantity is large, i.e.\ $\beta \gg (2\ell+1)$, we can neglect both $(2\ell+1)^2$ and the term linear in $\beta$
in the square-root, and also the term $(-1)$ outside, giving
\begin{equation} \label{eq:StrongFieldExpansion}
\Delta E_{m}^\pm = B \muB \left(m\pm 1/2\right).
\end{equation}
The solutions to the corner elements given by Eq.~\eqref{eq:SolutionMaxM} also reduce to this expression if we ignore
$\ell$ compared to $(\ell+1)\beta$. This result is trivially obtained from Eq.~\eqref{eq:Delta_E_Perturb} if the
spin--orbit term $\vect{L}\cdot\vect{S}$ is ignored, and thus corresponds to the well-known expression for the normal
Zeeman effect.

In this strong-field limit, the states are split in $m$ but degenerate in $\ell$ (except of course for the dependence
on $\ell$ in the relativistic correction, Eq.~\eqref{eq:RelCorr}), as seen in Fig.~\ref{fig:zeeB20}. The eigenstates
are particularly simple, since the denominator in Eq.~\eqref{eq:angle} becomes very large, and so $\varphi \approx
\pi/4$. Since the levels tend to group, and because of the selection rule $\Delta m=0,\pm 1$ to be discussed shortly,
the spectrum for normal Zeeman effect of a transition $n \rightarrow n'$ will consist of three equidistant lines -- a
Lorentz triplet~\cite[p.~211]{BJ}.

In the opposite limit, where $\beta \ll (2\ell+1)$, we get
\begin{equation}\label{eq:WeakFieldExpansion}
\begin{split}
\Delta E_{m}^\pm &\approx
\frac{W}{2} \left[ 2m\beta-1 \pm (2\ell+1) \left\{ 1+\frac{1}{2}\frac{4m\beta}{(2\ell+1)^2} \right\} \right] \\[+2mm]
& = \left\{
\begin{array}{l}
 W \ell   +B\muB m \dfrac{2\ell+2}{2\ell+1} ,\\[+2mm]
-W(\ell+1)+B\muB m \dfrac{2\ell}{2\ell+1},
\end{array}
\right.
\end{split}
\end{equation}
if we neglect the term quadratic in $B$. The solutions to the corner elements are also incorporated in this formula
(the top expression), which is easily shown -- without approximations -- from Eq.~\eqref{eq:SolutionMaxM}, if we
realize that $m / (\ell+1/2)=\pm 1$.

We can identify the first term in Eq.~\eqref{eq:WeakFieldExpansion} as the contribution from the spin--orbit
interaction if the magnetic field is ignored. The two cases correspond to $j=\ell\pm 1/2$, where $j$ is the quantum
number associated with the total angular momentum $\vect{J}=\vect{L}+\vect{S}$ which is conserved in the absence of a
magnetic field. The energy contribution from the magnetic field can now be evaluated, within this approximation,
through the expression~\cite{BJ}
\begin{equation*}
\Delta E_{\textrm{magn}} (m_j) = \frac{\muB}{\hbar} B \Big\langle L_z+2S_z \Big\rangle = \frac{\muB}{\hbar} B
\Big\langle J_z+S_z \Big\rangle
\end{equation*}
(this is where we need $g_s=2$). The expectation value of $J_z$ is $\hbar m_j$ by definition, and one may show that
\begin{equation}
\braket{n\ell s j m_j}{S_z}{n\ell s j m_j} = (g-1)m_j\hbar
\end{equation}
where $s$ is the spin (i.e.\ 1/2 for an electron) and
\begin{equation}
g = 1+\frac{j(j+1)+s(s+1)-\ell(\ell+1)}{2j(j+1)}
\end{equation}
is the Land{\'e} factor. Thus $\Delta E_{\textrm{magn}} = g \muB B m_j$, which can be shown to give exactly the
remaining part of Eq.~\eqref{eq:WeakFieldExpansion}. We now also realize why the corner solutions ended up in the top
expression of Eq.~\eqref{eq:WeakFieldExpansion}, since $|m|=\ell+1/2$ only is allowed for the triplet state
$j=\ell+1/2$.

In the weak-field case the distance between two levels corresponding to subsequent values of $m$ within the same
multiplet will be the same for all levels (cf.\ Fig.~\ref{fig:zeeB2}). This case is, for historical reasons, called
anomalous Zeeman effect, but is actually the more commonly encountered one. Since the splitting of the levels for
different multiplets is different (i.e.\ dependent on $j$), there will be more lines in the spectrum than for the
normal Zeeman effect.

The approximate expressions for strong and weak fields can thus be derived as expansions of the exact result,
Eq.~\eqref{eq:FinalEnergy}. We may now also specify quantitatively under what conditions these approximations hold,
since this is determined by whether $\beta=B/B_0=B\muB/W$ is much greater or much smaller than $2\ell+1$. Inserting the
constants into Eq.~\eqref{eq:WB0} gives
\begin{equation*}
\begin{split}
W &= \frac{Z^4}{\gamma} \cdot
3.62261\cdot10^{-4} \; \textrm{eV} = \frac{Z^4}{\gamma} \cdot 2.92183 \; \textrm{cm}^{-1}, \\[+2mm]
B_0 &= \frac{Z^4}{\gamma} \cdot 6.2584 \textrm{ Tesla}
\end{split}
\end{equation*}
where $\gamma = n^3 \ell (\ell+1/2)(\ell+1)$. Thus the approximations are valid if $B$ is much larger or smaller than
\begin{equation}\label{eq:StrongWeakCriterion}
\frac{(Z\alpha)^4}{n^3 \ell(\ell+1)} \frac{m_\textrm{e}c^2}{2\muB} = \frac{Z^4}{n^3 \ell(\ell+1)} \cdot 12.5 \textrm{
Tesla}.
\end{equation}
For light atoms and states with $n$ and $\ell$ less than 4, this expression takes values between 0.1 and 10 Tesla; for
heavier atoms the fourth power of $Z$ will dominate and, except for high shells and strong fields, the weak-field
approximation will almost always be valid.

As an example, let us consider the states $n=3$ and 4 in HeII ($Z=2$). Inserting $\ell=1$ and 2 for $n=3$ (for $\ell=0$
the strong-field approximation is exact; see the following section) the threshold field strengths come out as 3.70 and
1.23~Tesla respectively. For $n$=4 the corresponding values for $\ell$=1, 2 and 3 become 1.56, 0.52 and 0.26~Tesla. As
briefly mentioned in the Introduction, a modern application of Zeeman effect is to use fusion plasma spectroscopy in
order to determine the plasma temperature, by e.g.\ studying the transition between precisely these states in
HeII~\cite{Hellermann}. Typical magnetic fields in a tokamak the size of JET's lie in the range 1--3~Tesla, and we thus
clearly see that such an investigation cannot be carried out using either of the traditional approximations.


\section{Treatment of s-states} \label{sec:s-states}

As mentioned above, the expectation value $W$ is only defined for $\ell\neq0$ (cf.\ Eq.~\eqref{eq:Wnl}). But, in light
of the previous section, we now realize that $\ell=0$ corresponds to an extreme case of
Eq.~\eqref{eq:StrongFieldExpansion}, where the magnetic field is infinitely much stronger than the spin--orbit
interaction. Hence, without approximation, the energy eigenvalues are (here we may keep a general value of $g_s$)
\begin{equation}
\Delta E_{m}^\pm = B \muB \left(m \pm \frac{g_s-1}{2}\right) = \pm B \muB g_s/2, \qquad (\ell=0),
\end{equation}
since for $\ell=0$ we have $m=m_s=\pm 1/2$. To this we should also add the Darwin term~\cite{BJ}
\begin{equation}
\Delta E_{\textrm{D}} = \frac{\pi\hbar^2}{2m_{\textrm{e}}^2c^2} \frac{Ze^2}{4\pi\epsilon_0}
\braket{n00}{\delta(r)}{n00} = \frac{m_{\textrm{e}}c^2}{2}\frac{(Z\alpha)^4}{n^3}.
\end{equation}


\section{Transitions and matrix elements} \label{sec:HTrans}

To calculate the Zeeman pattern of the spectral line from a transition between two different $n\ell$ states, we need to
consider the selection rules and the transition matrix elements. Since the eigenstates (\ref{eq:ZeemanEigenstates}b) of
the total Hamiltonian, Eq.~\eqref{eq:TotalHamiltonian}, are simply linear combinations of hydrogenic states $\ket{n\ell
m_\ell m_s}$, it is however enough to employ the well-known results for transitions between two hydrogenic states
(derived in the absence of any magnetic field).

The derivation of the matrix elements for spontaneous emission is a procedure found in any textbook on the theory of
atomic spectra (see e.g.\ Ref.~\citealp{BJ}, Chapter~4). We therefore merely state the results in order that this paper
can be used as a self-contained reference on how to calculate the Zeeman spectrum. We will then give an illustrative
example of the resulting spectrum when these expressions are applied to the Zeeman problem at hand.


\subsection{Spontaneous emission matrix elements}

In the dipole approximation, the transition probability for spontaneous emission from a state $b$ to a lower state $a$
is found from Fermi's Golden Rule to be
\begin{equation} \label{eq:FermiGoldenRule}
W_{ab}^{\textrm{s}} (\Omega)\ d\Omega = \mathcal{C}(\omega_{ba}) \left| \uvecti{\epsilon} \cdot \vect{r}_{ba}
\right|^2\ d\Omega ,
\end{equation}
where $d\Omega$ is the solid angle element in which the radiation is observed, and $\uvecti{\epsilon}$ is a unit vector
in the polarization direction of the radiation. The overlap matrix element
\begin{equation}
\vect{r}_{ba} = \braket{n'\ell' m'_\ell m'_s}{\vect{r}}{n\ell m_\ell m_s} = \braket{n'\ell' m'_\ell}{\vect{r}}{n\ell
m_\ell} \delta_{m'_s m_s},
\end{equation}
where we introduced the prime notation for the higher state $b$ for a cleaner notation, and used the fact that the
operator $\vect{r}$ does not act on the spin space. Hence we obtain the first selection rule, $\Delta m_s=0$.

The pre-factor in Eq.~\eqref{eq:FermiGoldenRule} is
\begin{equation}
\mathcal{C}(\omega_{ba}) = \left( \frac{2Z}{a_\mu} \right) ^2 \frac{(\Delta E)^3}{\hbar (\mu c^2)^2} \frac{1}{8\pi Z^2
\alpha}
\end{equation}
where $\Delta E = \hbar\omega_{ba} = E_b-E_a$. It is generally only meaningful to compare the relative intensities of
Zeeman components for the same chemical element (same $Z$ and $\mu$), and hence the only factor we really need to
retain in $\mathcal{C}$ is the energy difference $\Delta E$. When calculating the spectral line intensity from the
transition probability, an extra factor $\Delta E$ enters, making the observed intensity proportional to $(\Delta
E)^4$.

Introducing the spherical tensor components of $\uvecti{\epsilon}$
\begin{equation}
\epsilon_0 = \uvecti{\epsilon}_z \qquad \epsilon_{\pm 1} = \mp \frac{1}{\sqrt{2}} (\uvecti{\epsilon}_x \pm i
\uvecti{\epsilon}_y) ,
\end{equation}
and correspondingly for $\vect{r}_{ba}$, we can write
\begin{equation} \label{eq:epsscalarr}
\uvecti{\epsilon} \cdot \vect{r}_{ba} = \sum_{q=-1}^1 \epsilon_q^* \vect{r}_{ba}^q = \sum_{q=-1}^1 \epsilon_q^*
\mathcal{I}_{n' \ell' m_\ell' n \ell m_\ell}^q
\end{equation}
where
\begin{equation} \label{eq:Idef}
\mathcal{I}_{n' \ell' m_\ell' n \ell m_\ell}^q = \sqrt{\frac{4\pi}{3}}\ \mathcal{R}_{n\ell}^{n'\ell'} \int_\Omega
Y_{\ell' m_\ell'}^* Y_{1q} Y_{\ell m_\ell} d\Omega .
\end{equation}
with the radial overlap integral
\begin{equation} \label{eq:Rint}
\mathcal{R}_{n\ell}^{n'\ell'} = \int_0^\infty R_{n'\ell'}(r)\ R_{n\ell}(r)\ r^3\ dr .
\end{equation}
Here $R_{n\ell}(r)$ is the radial part of the hydrogenic wavefunction, to which we will shortly return.

Using either the Wigner--Eckart theorem or the properties of the spherical harmonics, one can show that~\cite[Section
11-4 and Chapter 14]{Cowan}
\begin{equation} \label{eq:intensfinal}
\mathcal{I}_{n' \ell' m_\ell' n \ell m_\ell}^q = (-1)^{\ell_>-m_\ell'} \sqrt{\ell_>} \mathcal{R}_{n\ell}^{n'\ell'}
\begin{pmatrix} \ell & 1 & \ell' \\[+2mm] m_\ell & q & -m_\ell' \end{pmatrix},
\end{equation}
where $\ell_> = \textrm{max}(\ell',\ell)$. The properties of the Wigner 3j-symbol now enable us to identify further
selection rules, since the 3j-symbol vanishes unless $\ell'=\ell+1$ and $m_\ell+q=m_\ell'$. Thus we obtain the $\ell$
selection rule which requires change of parity; dipole transitions only connect states of opposite parity, and the
parity of a hydrogenic state is $(-1)^\ell$. Moreover, since $q=0,\pm 1$ we may write the other selection rule as
$\Delta m_\ell=0,\pm 1$. In summary, the dipole selection rules for hydrogenic states are
\begin{equation}
\Delta m_s=0, \qquad \Delta \ell=1, \qquad \Delta m_\ell=0,\pm 1.
\end{equation}


\subsection{The radial integral} \label{sec:radial}

Including proper normalization, the radial part of the hydrogenic wavefunction becomes
\begin{equation} \label{eq:Rnl}
R_{n\ell}(r) = \sqrt{ \left( \frac{2Z}{a_\mu} \right)^3 \frac{(n-\ell-1)!}{2n^4(n+\ell)!}}\ e^{\rho/2n} \left(
\frac{\rho}{n} \right)^\ell\ \mathcal{L}_{n-\ell-1}^{2\ell+1}(\rho/n) ,
\end{equation}
where $\mathcal{L}_{n-\ell-1}^{2\ell+1}(x)$ are the associated Laguerre polynomials, $a_\mu = 4\pi\epsilon_0
\hbar^2/\mu e^2$ is the Bohr radius (adjusted for finite nuclear mass), and we have also introduced $\rho = 2Zr/a_\mu$
as the dimensionless length scale.

Inserting this into the integral~\eqref{eq:Rint}, one arrives at
\begin{equation}\label{eq:MatrixElementFinal}
\mathcal{R}_{n\ell}^{n'\ell'} = \frac{a_\mu}{2Z} \mathcal{F}_{n\ell}^{n'\ell'}
\end{equation}
with the dimensionless form factor
\begin{equation}
\begin{split}
\mathcal{F}_{n\ell}^{n'\ell'} &= \frac{1}{2} \sqrt{\frac{(n-\ell-1)!\ (n'-\ell'-1)!}{(n+\ell)!\ (n'+\ell')!}}
\frac{1}{n^{\ell+2}} \frac{1}{{n'}^{\ell'+2}} \\[+1mm]%
&\quad \times \int_0^\infty e^{-\frac{\rho}{2}\left( \frac{1}{n}+\frac{1}{n'} \right)}\ \rho^{\ell+\ell'+3}\
\mathcal{L}_{n-\ell-1}^{2\ell+1} (\rho/n)\ \mathcal{L}_{n'-\ell'-1}^{2\ell'+1} (\rho/n')\ d\rho.
\end{split}
\end{equation}


\subsection{Polarization}

\begin{figure}[hbt]
\centering \epsfig{file=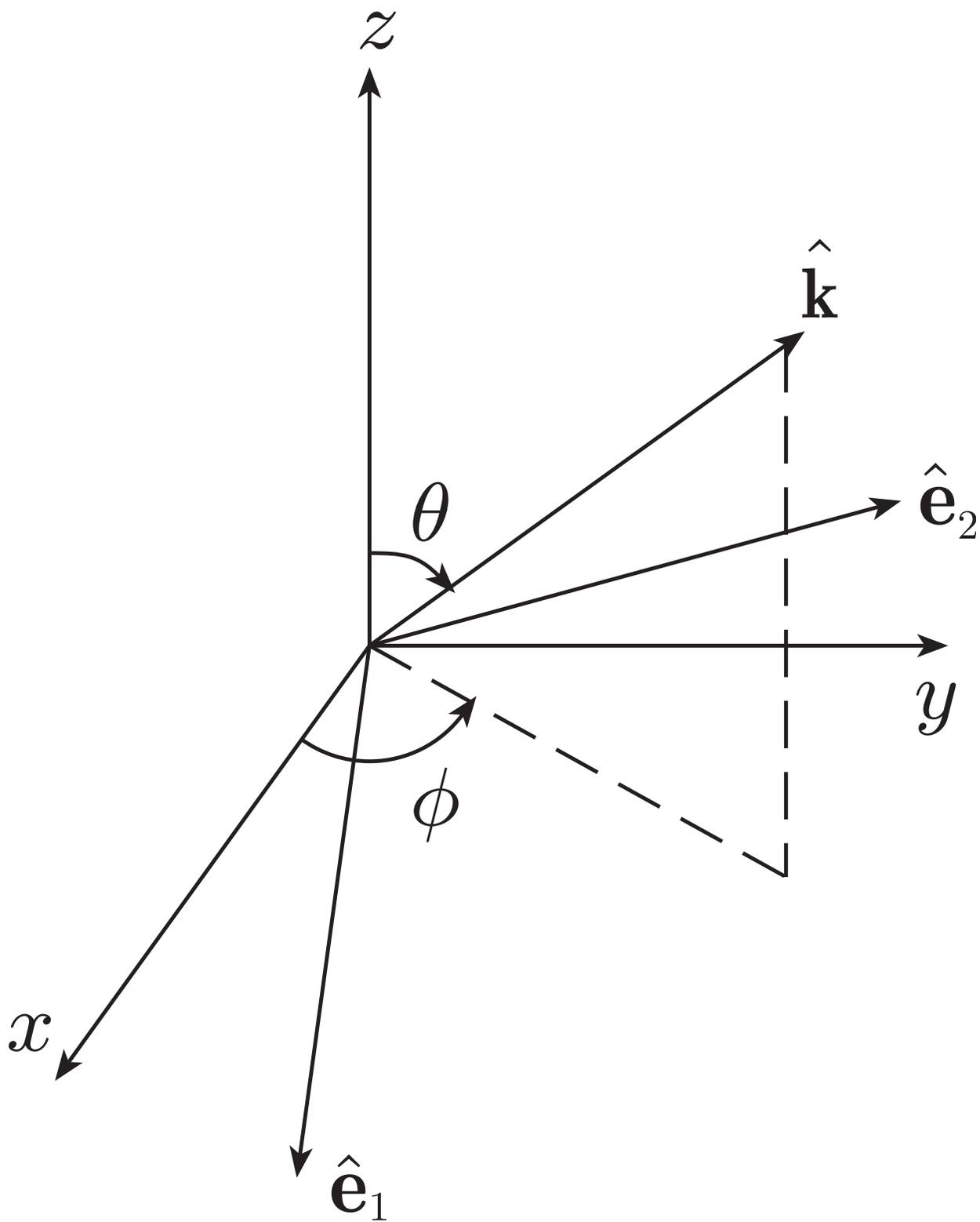,width=\linewidth} \caption{Definition of the polarization vectors.}
\label{fig:polvect}
\end{figure}

We now return to Eqs.~\eqref{eq:FermiGoldenRule} and \eqref{eq:epsscalarr} to study the polarization of the emitted
radiation. Any arbitrary polarization can be expressed as a linear combination of two independent directions
$\uvect{e}_1$ and $\uvect{e}_2$, such as left and right circularly polarized, or two perpendicular linear
polarizations. For this discussion we choose the latter. We also introduce a wavevector $\uvect{k}$ of unit length (as
are the $\vect{e}$-vectors) pointing in the direction of propagation of the emitted light, so that the three vectors
$\uvect{e}_1$, $\uvect{e}_2$ and $\uvect{k}$ form a right-handed system (see Fig.~\ref{fig:polvect}). As before we let
the $z$-axis be the direction of the magnetic field. Taking, arbitrarily, $\uvect{e}_2$ to lie in the $xy$-plane and
$\uvect{e}_1$ pointing downwards, we have \addtocounter{equation}{1}
\begin{align}
\tag{\theequation a} \uvect{k}    &= (\sin\theta\cos\phi,\sin\theta\sin\phi,\cos\theta), \\[+1mm]
\tag{\theequation b} \uvect{e}_1 &= (\cos\theta\cos\phi,\cos\theta\sin\phi,-\sin\theta), \\[+1mm]
\tag{\theequation c} \uvect{e}_2 &= (-\sin\phi,\cos\phi,0).
\end{align}

Next we use the properties of the 3j-symbol in Eq.~\eqref{eq:intensfinal}, which is non-zero if and only if
$m_\ell'=m_\ell+q$, as already stated. Since each of the three values of $q$=0 or $\pm1$ only may be satisfied one at a
time, we have three separate cases to study. These are for historical reasons referred to as $\pi$ and $\sigma^\pm$
(from the German words for parallel and perpendicular, "senkrecht").

\begin{itemize}

\item[$\boldsymbol{\pi}$]
In this case $\Delta m_\ell=0$, and consequently we only need to consider $\mathcal{I}^0$ (we suppress the quantum
numbers in this part). This gives the transition rate
\begin{equation}
W_{ab}^{\textrm{s}} (\pi) = \mathcal{C}(\omega_{ba}) \sin^2\theta \left| \mathcal{I}^0 \right|^2
\end{equation}
for a photon with polarization $\uvect{e}_1$, whereas the rate is zero for polarization $\uvect{e}_2$. Since the
$\uvect{k}$-vector is defined as the direction in which the radiation is observed, we find that if the emission is
viewed longitudinally ($\uvect{k}$ parallel to the magnetic field), $\theta=0$ and so the $\pi$ component is absent. In
the transverse direction $(\theta=\pi/2)$, i.e. in a direction perpendicular to the magnetic field, we have
$\uvect{e}_1=(0,0,-1)$ and thus the $\pi$ component will be plane polarized with $\uvecti{\epsilon}=\uvect{e}_1$ in the
direction of the negative $z$-axis.

\item[$\boldsymbol{\sigma^+}$]
This case corresponds to $\Delta m_\ell = m'_\ell-m_\ell=-1$. Now the only contribution will be from
$\mathcal{I}^{-1}$, and since the two polarization directions are orthogonal and independent, we may sum the
probabilities instead of the amplitudes. In result, we obtain
\begin{equation}
W_{ab}^{\textrm{s}}(\sigma^+) = \mathcal{C}(\omega_{ba}) \frac{1+\cos^2\theta}{2} \left| \mathcal{I}^{-1} \right|^2.
\end{equation}
The $\sigma^+$ component is plane polarized with $\uvecti{\epsilon}=\uvect{e}_2$ in transverse observation, whereas in
longitudinal observation it is circularly (left-hand) polarized. The intensity of each $\sigma$ component is always
half of the $\pi$ component in transverse observation.

\item[$\boldsymbol{\sigma^-}$]
For the case $\Delta m_\ell=1$ we get the analogous expression
\begin{equation}
W_{ab}^{\textrm{s}}(\sigma^-) = \mathcal{C}(\omega_{ba}) \frac{1+\cos^2\theta}{2} \left| \mathcal{I}^{+1} \right|^2 .
\end{equation}
The $\sigma^-$ component is also plane polarized in transverse observation, and (right-hand) circularly polarized in
longitudinal.

\end{itemize}

\begin{figure}[t]
\centering \epsfig{file=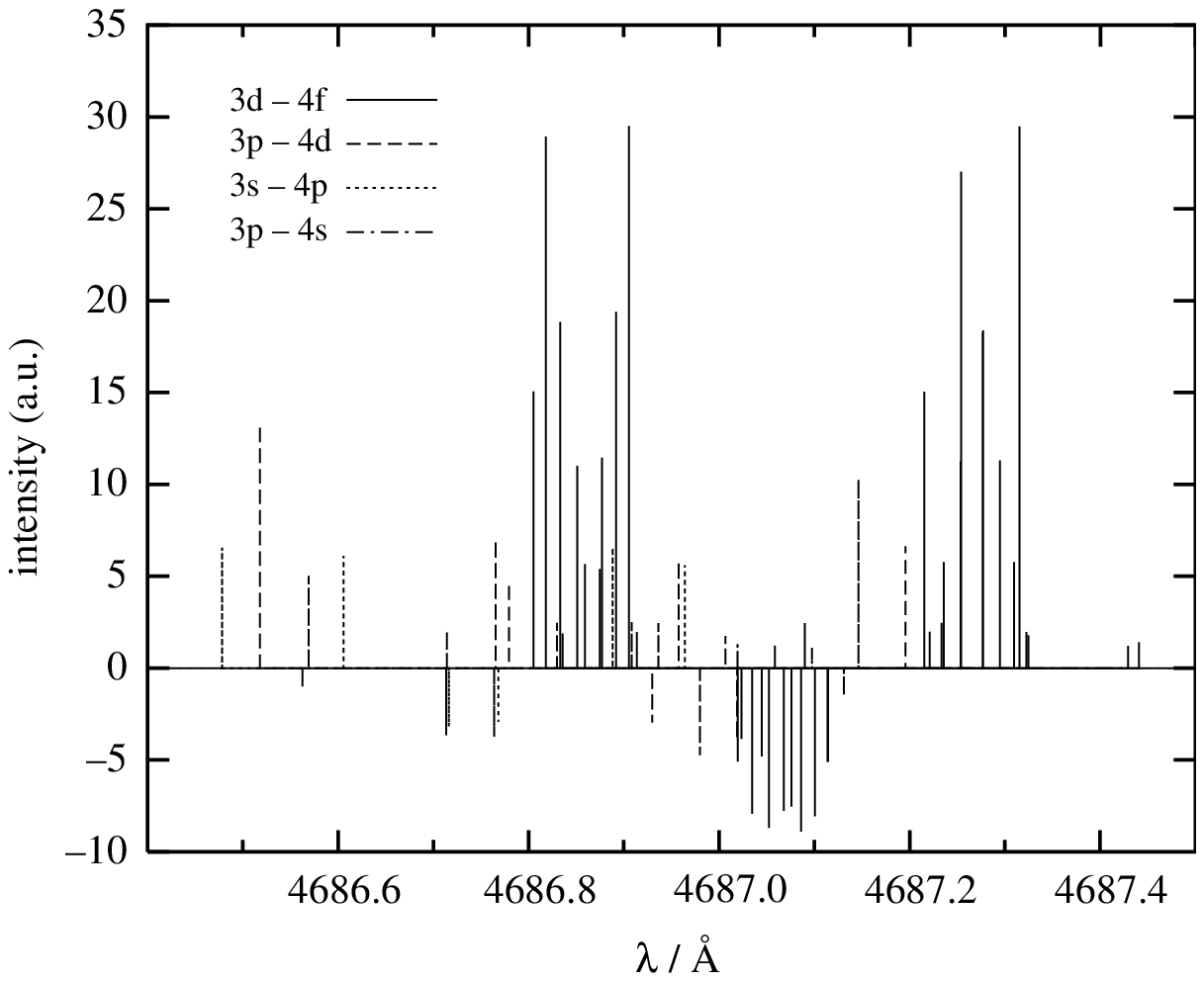, width=\linewidth} \caption{The complete Zeeman pattern for the HeII line at
4687~\AA\ (vacuum) consists of 146 components. It is here calculated for $B=2$~Tesla at a viewing angle
$\theta=40\degree$ relative to the magnetic field axis (cf. Fig.~\ref{fig:polvect}). Only components with intensity
greater than one (in the plot units) are included; the $\pi$ components are conventionally drawn with negative
intensity.} \label{fig:ZeeHeII}
\end{figure}

From here on it is a simple matter to calculate the intensities of all transitions in the Zeeman pattern; see
Fig.~\ref{fig:ZeeHeII} for an example. The number of components increases rapidly with the quantum numbers; for a $7-8$
transition the total number is 512, although many will be very weak.


\section{Discussion of approximations} \label{sec:Approxs}

In this section we summarize the approximations made in the treatment. They are not many, as we have strived to keep
things as general and exact as possible. For an even more detailed discussion, see Ref.~\citealp{BS}, Sections
47$\alpha$--$\epsilon$.

\begin{itemize}
\item
{\bf The quadratic term} \\
In Section~\ref{sec:ExternalField} we ignored the term in the Hamiltonian quadratic in $\vect{A}$, and we also showed
that for fields of normal strengths (less than 1000 Tesla or so), this approximation is certainly valid, and simplifies
the algebra considerably.

A particular effect which enters if the quadratic term is retained is the possibility of mixing states with different
$\ell$ quantum numbers. The linear term in $\vect{A}$ commutes with $\vect{L}^2$, making $\ell$ a good quantum number,
but the quadratic term does not. Specifically, it introduces mixing between states with $\ell$-values differing by $\pm
2$. Again, this will only be of importance for very strong magnetic fields and can often safely be ignored.

\item
{\bf Relativistic corrections}\\
Using the Pauli approximation instead of the Dirac theory means we are only calculating the Zeeman effect to lowest
order in $\alpha^2$. Assuming that all perturbations (magnetic field and spin--orbit interaction) are small, we remedy
this by using the same correction Eq.~\eqref{eq:RelCorr} as for the field-free case to correct for the non-relativistic
momentum operator. Effects beyond this should be negligible.

\item
{\bf The anomalous magnetic moment}\\
In the approximate treatment of the anomalous Zeeman effect (weak fields) it is necessary to use the value $g_s=2$. As
it turned out, we were not restricted by this approximation in our exact derivations (note the factors $g_s-1$ and
$g_s/2$ appearing in the final expressions, Eqs.~\eqref{eq:SolutionMaxM} and \eqref{eq:FinalEnergy}), although the
influence is very small -- unless of course $g_s$ differs significantly from 2.

\item
{\bf Nuclear motion and hyperfine structure}\\
We have already taken into account the nuclear motion by using the reduced mass $\mu$ of the electron for the
central-field Hamiltonian. In the perturbations, we have kept the electron mass since the perturbations themselves are
small, so any corrections to them are even smaller~\cite{BJ}.

Hyperfine structure due to the nuclear magnetic moment is in general orders of magnitude smaller than the
fine-structure splitting (spin--orbit interaction). Therefore, we expect that hyperfine structure will not perturb the
Zeeman spectrum except for extremely weak magnetic fields, where the Zeeman pattern can be heavily influenced.

\end{itemize}


\section{Summary} \label{sec:Summary}

In conclusion, this paper has demonstrated how the problem of a single-electron atom or ion in a static magnetic field
can be solved exactly, under the only simplifying approximation that the term which is quadratic in the field strength
can be neglected in the Hamiltonian. No assumption was made regarding the value of the electron spin gyromagnetic
ratio, which allows the obtained results to be readily generalized to applications in e.g.\ solids.

By using the results presented here, a unified treatment of the Zeeman effect becomes possible over the entire range of
fields presently employed in e.g.\ fusion plasma, where the influence of the Zeeman effect on the plasma temperature
measurements has been demonstrated to be significant in many cases. The well-known formulas for the weak and strong
field limits are easily obtained as expansions of the general results, and in addition it was possible to obtain a
quantitative expression for the fields at which these approximations are valid.

The obtained general expressions for the eigenstates and their corresponding energies are not very complicated. As a
final remark, one may therefore point out that the derivation, as presented here, has certain pedagogical merits and
provides a good example of how to apply the power of matrix algebra to quantum-mechanical problems.


\section*{Acknowledgement}

The author wishes to thank I. Martinson and C. Jup\'en for many useful discussions related to the work presented here.


\begin{thebibliography}{99}

\bibitem{Zeeman}
Zeeman, P., Philosophical Magazine {\bf 43}, 226 (1897).

\bibitem{X}
White, H.~E., "Introduction to Atomic Spectra," (McGraw-Hill Book Co., New York, 1934), Chapter~X.

\bibitem{Darwin}
Darwin, C.~G., Proc. Royal Soc. London {\bf 115}, 1 (1927).

\bibitem{bethe}
Bethe, H., in  "Handbuch der Physik," (edited by Smekal, A.) (Springer Verlag, Berlin, 2nd edition, 1933), vol. XXIV,
part 1, pp.~396--399.

\bibitem{BS}
Bethe, H.~A. and Salpeter, E.~E., "Quantum Mechanics of One- and Two-Electron Systems" (Springer Verlag, Berlin, 1957).

\bibitem{BJ}
Bransden, B.~H. and Joachain, C.~J., "Physics of Atoms and Molecules" (Longman Scientific \& Technical, Harlow, 1983).

\bibitem{Islerreview}
Isler, R.~C., Plasma Phys. Control. Fusion {\bf 36}, 171 (1994).

\bibitem{ppcf}
Blom, A. and Jup\'en, C., Plasma Phys. Control. Fusion {\bf 44}, 1229 (2002).

\bibitem{hey}
Hey, J.~D., Lie, Y.~T., Rusb{\"u}ldt, D. and Hintz, E., Contrib. Plasma Phys. {\bf 34}, 725 (1994).

\bibitem{Fano}
Fano, U. and Fano, L., "Physics of Atoms and Molecules. An Introduction to the Structure of Matter" (University of
Chicago Press, Chicago, 1972).

\bibitem{simola}
Simola, J. and Virtamo, J., J. Phys. B: Atom. Molec. Phys. {\bf 11}, 3309 (1978).

\bibitem{callaway}
Callaway, J., "Quantum Theory of the Solid State," (Academic Press, San Diego, 2nd edition, 1991), pp.~526--529.

\bibitem{Hellermann}
von Hellermann, M. {\em et al.}, Plasma Phys. Control. Fusion {\bf 37}, 71 (1995).

\bibitem{Cowan}
Cowan, R.~D., "The Theory of Atomic Structure and Spectra" (University of California Press, Berkeley, 1981).

\end{thebibliography}
\end{document}